\documentclass[aps,prl,reprint,groupedaddress]{revtex4-2}

\usepackage{graphicx,amsmath,color,amsfonts, amsmath, bm}

\usepackage{comment}
\usepackage[dvipsnames]{xcolor}
\newcommand{\updt}[1]{#1}

\begin{document}

\title{Phase Separation Driven by Active Flows}

\author{Saraswat Bhattacharyya, Julia M.\ Yeomans}
\affiliation{Rudolf Peierls Centre For Theoretical Physics, University of Oxford, Oxford OX1 3PU, United Kingdom.}

\date{\today}

\begin{abstract}

We extend the continuum theories of active nematohydrodynamics to model a two-fluid mixture with separate velocity fields for each fluid component, coupled through a viscous drag.
The model is used to study an active nematic fluid, mixed with an isotropic fluid. We find micro-phase separation, and argue that this results from an interplay between active anchoring and active flows driven by concentration gradients. The results may be relevant to cell-sorting and the formation of lipid rafts in cell membranes.

\end{abstract}

%\pacs{}
%\keywords{}


\maketitle

\begin{figure}[ht]
    \centering
   \includegraphics[scale=0.88]{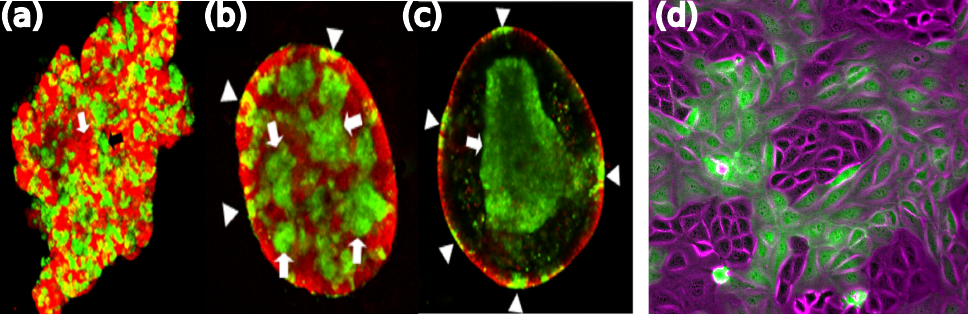}
    \caption{(color online.) {\bf Cell Sorting}: (a), (b), (c) starfish embryo at times (a) t= 4 hr, (b) t= 8 hr, (c) t = 48 hr showing sorting of embryonic cells (adapted from Ref \cite{Suzuki2021}). (d) Phase ordering of extensile (magenta) and contractile (green) epithelial MDCK cells (courtesy of L. Balasubramaniam \cite{Balasubramaniam2021}).}
%   hradapted from Ref \cite{Suzuki2021}, show two different cell types (marked by green and red) sorting themselves over time in a starfish embryo. Time snapshots are (a) t= 4 hr, (b) t= 8 hr, and (c) t = 48 hr. Panel (d) shows cell sorting in a mixture of epithelial MDCK cells, where the magenta cells are extensile and the green cells are contractile, obtained from L. Balasubramaniam's lab. Cell sorting is seen in starfish embryos, hydra and epithelial co-cultures  \cite{GOMEZGALVEZ2021, Suzuki2021, SKOKAN2020, Skamrahl2022, Balasubramaniam2021}}
    \label{fig:CellSorting}
\end{figure}

{\bf Introduction.}---There is increasing evidence that phase ordering plays a fundamental role in biological processes such as morphogenesis and colony growth  \cite{Halbleib2006, KRENS2011, Durand2021, Skamrahl2022}.  Phase ordering of cell types (Fig.~\ref{fig:CellSorting})  \cite{STEINBERG1975}, membrane-less organelles \cite{Banani2017} and RNA-protein mixtures \cite{Dutagaci2021} inside cells have been studied using free-energy prescriptions. However, biological matter is intrinsically out of thermodynamic equilibrium, and it is unlikely that biological phase ordering relies entirely on the framework of equilibrium thermodynamics.

Self-motile particles form an important class of non-equilibrium systems called active matter \cite{Ramaswamy2010}. Active matter exhibits  non-equilibrium phase segregation mechanisms, such as motility-induced phase separation (MIPS) \cite{Cates2015, Gonnella2015, Stenhammer2013} where active particles get trapped in regions of high particle density. Phase separation has been found in continuum models \cite{Wittkowski2014, Redner2013, Kuan2021}, in the presence of hydrodynamic interactions \cite{Thutupalli2018}, or driven by aligning torques \cite{Zhang2021, Gokhale2022}.  
Active nematics consist of active rod-like particles with orientational order. The elongated particles generate active dipolar stresses which destabilise the aligned state and lead to active turbulence characterised by a chaotic and highly vortical velocity field
\cite{Doostmohammadi2018, Marchetti2013, Duclos2017}. Recent work has shown that bacterial biofilms \cite{Yaman2019, DellArciprete2018}, epithelial tissues \cite{Saw2017, BALASUBRAMANIAM2022}, and microtubule-motor mixtures \cite{sanchez2012} can be modelled as active nematics. There is little work about phase ordering in active nematics \cite{Blow2014, Caballero2022}, but recently microphase separation was predicted to occur in extensile gels near the nematic-isotropic transition point \cite{Assante2022}. 
%\sar{Active stirring in phase separating mixtures leads to a steady state characterized by finite domain lengths \cite{Blow2014}.} 

It is still not clear whether and how active stresses can lead to phase ordering in a binary mixture of an active nematic and a passive fluid in the absence of any thermodynamic driving force. To answer this question, we extend the continuum theories of active nematohydrodynamics to a two-fluid model  with separate velocity fields, coupled through viscous drag, for each fluid component \cite{stewart1984, levy1999}. This allows us to model relative motion between the two fluids which is a requirement for activity-driven phase ordering. Two-fluid models that have been used in biological contexts include studies of cytoskeleton dynamics and growing biofilms \cite{Mogilner2018, Cogan2010, Alt1999}.

{\bf Model.}---We model a binary mixture of an active nematic fluid (component 1), and a passive isotropic fluid (component 2), in $d$ dimensions. Each fluid component, denoted by $i \in {1, 2}$, has a density field $\rho^{i}$, velocity field $\mathbf{u}^{i}$ and a chemical potential $\mathbf{\mu}^{i}$. 
The active nematic is associated with a second rank tensor order parameter field \cite{degennes_book},  $
    \mathbf{Q} =  \frac{d}{d-1} S \big( \mathbf{n n} - \frac{ \mathbb{I}}{d} \big)
$, where \updt{$S$ denotes the degree of nematic ordering, and} $\mathbf{n}$ denotes the orientation of the local nematic director field.

Each component fluid $i$ obeys the mass continuity equation and the momentum balance equations
\begin{align}
    \partial_t \rho^{i} + \nabla \cdot( \rho^{i} \mathbf{u}^{i} ) = &0, \label{eqn:CompMassContinuity} \\
    \partial_t ( \rho^{i} \mathbf{u}^{i}) + \nabla \cdot ( \rho^{i} \mathbf{u}^{i} \mathbf{u}^{i} )= &-\rho^{i} \nabla \mu^{i} + \gamma \phi (1-\phi) (\mathbf{u}^{3-i} - \mathbf{u}^{i}) \nonumber \\ &+ \nabla \cdot \mathbf{\sigma}^{visc, i} + \delta_{i1}\nabla \cdot \mathbf{\sigma}^{nem},
    \label{eqn:CompMomentumBalance}
\end{align}
where $\phi = \rho^1/\rho^c$ is the concentration of the active fluid, and $\rho^c = \rho^1 + \rho^2$ is the total fluid density.
The forces acting on the fluids, \updt{on the rhs of  Eq.~(\ref{eqn:CompMomentumBalance})}, are thermodynamic forces, viscous drag between the component fluids, internal viscous dissipation \updt{$\sigma^{visc, i} = \eta^i \big(\, (\nabla \mathbf u^i) + (\nabla \mathbf  u^i)^T - \frac{2}{d} (\nabla\cdot \mathbf u^i) \mathbb{I} \,\big)$}, and, for the active nematic fluid,  elastic and active forces \cite{Doostmohammadi2018}, \updt{respectively. The stresses in the nematic are} given by 
\begin{align} 
\allowdisplaybreaks
 \mathbf{\sigma}^{el} &= 2\lambda \big(\mathbf{Q} + \mathbb{I}/d \big) (\mathbf{Q : H})-  \lambda \mathbf{H}\cdot \big( \mathbf{Q} + \mathbb{I}/d \big)- \label{eqn:elasticStress}   \\
& \lambda \big( \mathbf{Q}  + \mathbb{I}/d \big)\cdot\mathbf{H} -\big( \nabla \mathbf{Q} \big)\, . \frac{\partial \mathcal F}{\partial \nabla \mathbf{Q}}  + \mathbf{Q \cdot H} - \mathbf{H \cdot Q},  \nonumber \\
  \mathbf{\sigma}^{act}  &= -\zeta \mathbf{Q} ,\label{eqn:activeStress}  \\
    \mathbf{\sigma}^{nem} &= \phi \, \big(\, \mathbf{\sigma}^{el} + \mathbf{\sigma}^{act} \big),  \label{eqn:totalStress} 
\end{align}
where $\mathcal F$ is the free energy, $\mathbf{H}$ is the molecular field \cite{Doostmohammadi2018} defined  by 
$
    \mathbf{H} =-\frac{\partial \mathcal{F}}{\partial \mathbf{Q}} + \frac{\mathbb{I}}{d} \text{Tr}\bigg( \frac{\partial \mathcal{F}}{\partial \mathbf{Q} } \bigg)   
$, $\lambda$ is the flow aligning parameter, and $\zeta$ is the magnitude of the activity. \updt{Eq. (\ref{eqn:elasticStress}) is the (passive) elastic stress \cite{degennes_book}, while Eq. (\ref{eqn:activeStress}) is the active stress which produces pusher ($\zeta>0)$ or puller ($\zeta<0$) dipolar flows.}

The nematic tensor field evolves according to \cite{Beris1994}
\begin{flalign}
    \partial_t \mathbf{Q} + \mathbf{u}^1 \cdot \nabla \mathbf{Q} - \mathbf{S} = \Gamma \mathbf{H}
    \label{eqn:Q_Field}
\end{flalign}
where
$
    \mathbf{S} = (\lambda \mathbf{E}^1 + \mathbf{\Omega}^1)\cdot \big( \mathbf{Q}+\mathbb{I}/d \big) + 
    \big( \mathbf{Q}+\mathbb{I}/d \big)\cdot (\lambda \mathbf{E}^1 - \mathbf{\Omega}^1) %\nonumber\\&
     -  2\lambda \big( \mathbf{Q}+\mathbb{I}/d \big) (\mathbf{Q} :\nabla \mathbf{u}^1)
    %\label{eqn:Corot_term}
$
is the co-rotation term \updt{describing the response of the orientation field to strain and vorticity in the flow, while the right hand side describes the relaxation to a state of minimum free energy. Here,} $\mathbf{E}^1$ the rate of strain tensor and $\mathbf{\Omega}^1$ the vorticity tensor of the active fluid.

Equilibrium is described by a free energy \cite{Doostmohammadi2018, Marchetti2013, Blow2014, Malevanets1999}
\begin{align}
    \mathcal{F} = \int d^2\mathbf{r}  \bigg[ \rho^c \Big( \frac{1}{3} \ln \rho^c + a \,(\phi - 1/2)^2 \Big) + \frac{1}{2} \kappa || \mathbf{\nabla}\phi ||^2\nonumber \\  + 
     \frac{C}{2} \big( S_{nem}^2 \phi - \frac{1}{2} \text{Tr}(\mathbf{Q}^2) \big)^2 + \frac{K_{LC}}{2} ( \nabla \mathbf{Q} )^2  \bigg]
    \label{eqn:FreeEnergy}
\end{align}
%where $C$ is the nematic bulk energy constant, $K_{LC}$ is the nematic elastic constant, $a$ is the coefficient of the restoring free energy, and $\kappa$ is the coefficient of surface tension. %
where the nematic elastic constant $K_{LC}$, $C$, $a$ and $\kappa$ are material parameters. \updt{The first term represents the pressure contribution to the free energy from the isothermal equation of state for the fluid \cite{LB_book}. The rest of the first line describes a Landau-Ginzburg free energy with $a>0$ favouring a homogeneous mixed state ($\phi=1/2$) and $\kappa>0$ penalizing interfaces. The second line is the Landau-de Gennes free energy describing an ordered nematic.}

We change variables to give one equation for the combined fluid, which is to a good approximation incompressible, and one for the active fluid. To do this, we define the centre of mass velocity of the combined fluid $ \mathbf{u}^c = \phi \mathbf{u}^1 + (1-\phi) \mathbf{u}^2 $, and the relative slipping velocity between the fluids $  \delta \mathbf{u} = \mathbf{u}^{1} - \mathbf{u}^{2} $. Assuming that the viscous drag between the components is the fastest relaxation process in the system, $\delta \mathbf{u}  << \mathbf{u}^1, \mathbf{u}^2$, we neglect all terms of order $(\delta \mathbf{u})^2$\updt{, and assume that the drag between the fluids is linear in $\delta \mathbf{u}$}. Moreover, recalling that $\mu^i = \partial \mathcal F / \partial \rho^i$,  the thermodynamic forces can be written as a stress tensor
%\begin{flalign}
   $ \sum_{i=1}^2 (-\rho^i \mathbf{\nabla} \mu^i) = -\mathbf{\nabla \cdot \sigma}^{th}
   $
%\end{flalign}
which, for the free energy in Eq.~(\ref{eqn:FreeEnergy}), is 
%\begin{align}
 $   \mathbf{\sigma}^{th} = p \mathbb{I} + \kappa \big( \mathbf{\nabla \nabla} \phi - \frac{1}{2} || \mathbf{\nabla} \phi ||^2 \mathbb{I} \big)
    \label{eqn:LG_stress} $
%\end{align}
where $p = \rho^c/3$ \cite{LB_book, Malevanets1999, Malevanets2000}.
% is an isotropic pressure term, and the second term describes an anisotropic stress due to surface tension \cite{LBbook, Malevanets1999, Malevanets2000}.
Assuming that both fluids have the same viscosity $\eta$,  the internal viscous dissipation of the combined fluid is 
%\begin{flalign}
    $\mathbf{F}^{visc, c} = \eta^c \nabla^2 \mathbf{u}^c$
%\end{flalign}
where $\eta^c = 2 \eta$.

Adding Eqs.~(\ref{eqn:CompMassContinuity}), (\ref{eqn:CompMomentumBalance}) for each component then shows that the combined fluid satisfies
\begin{align}
     & \partial_t \rho^{c} + \mathbf{\nabla} \cdot (\rho^{c} \mathbf{u}^{c}) = 0 \label{eqn:CombiMassContinuity}, \\
     & \partial_t (\rho^{c} \mathbf{u}^{c}) + \mathbf{\nabla}\cdot ( \rho^{c} \mathbf{u}^{c} \mathbf{u}^{c}) = -\mathbf{\nabla} \cdot \sigma^{th} + \mathbf{F}^{visc, c} + \mathbf{\nabla} \cdot \sigma^{nem} .
    \label{eqn:CombiMomentumBalance}
\end{align}
The equations for the first (active) component are
\begin{align}
     \partial_t \rho^{1} + \mathbf{\nabla} \cdot (\rho^{1} \mathbf{u}^{1}) = & \,0,
    \label{eqn:Fluid1MassContinuity} \\
     \rho^{1} [ \partial_t  \mathbf{u}^{1} + \mathbf{u}^{1}\cdot \mathbf{\nabla} \mathbf{u}^{1} ]  = &-\phi \mathbf{\nabla} \cdot \mathbf{\sigma}^{th} + \phi \mathbf{G} + \mathbf{F}^{visc, 1}\nonumber \\  &+ \mathbf{\nabla} \cdot \mathbf{\sigma}^{nem} 
    \label{eqn:Fluid1MomentumBalance}
\end{align}
where
%\begin{flalign}
$    \mathbf{G} = \gamma (\mathbf{u}^{c} - \mathbf{u}^{1})+ (1-\phi) \mathbf{\nabla} \delta \mu $
%\end{flalign}
is the force applied by the passive component, and $\delta \mu = -\delta \mathcal{F} /\delta \phi$. 

Eqs.~(\ref{eqn:CombiMassContinuity}-\ref{eqn:Fluid1MomentumBalance}) are solved using a lattice Boltzmann (LB) algorithm \cite{Malevanets1999, Malevanets2000}. For the equation of the $\mathbf{Q}$ field (\ref{eqn:Q_Field}) we use a finite difference (FD) approach. 5 FD steps are taken for each LB step to optimize simulation speed.
Simulations are run in a periodic box of size 200 x 200 for 20,000 LB time-steps, with a random initial director configuration. Parameter values are $\rho^1 = 20, \rho^2 = 20, \gamma=4, \eta_1 = \eta_2 = 10/3, a=0.0025, \kappa = 5, \Gamma = 0.1, C = 0.1, \zeta = 0.1, K_{LC} = 0.15, S_{nem} = 1, \lambda = 0$ unless otherwise indicated. 

%\noindent
{\bf Phase separation driven by activity.}---We solve the equations of motion for a mixture which comprises equal concentrations of an active nematic fluid and a passive fluid. The mixture is incompressible so that the total fluid density remains constant, but the relative concentrations of individual components can vary in space and time. The two components are initially homogeneously mixed so that the concentration of the active component $\phi = 0.5$ everywhere. However, above a threshold activity the system quickly self-organises into dynamically-evolving regions which are, respectively, rich or poor in the active fluid component.  A typical snapshot of the phase-ordered state is shown in Fig.~\ref{fig:ActivePhaseSeparation}(a), and Movie 1 in the Supplemental Material \updt{(SM)} \cite{supp} shows its time evolution. The domains continually break up and re-form. They are elongated by extensile active flows \cite{Blow2014} and then pulled apart by the chaotic active turbulent flows .

We measure the standard deviation of $\phi$, denoted by $\Delta$, to quantify the degree of phase separation. Fig.~\ref{fig:ActivePhaseSeparation}(b) shows how $\Delta$ evolves with time towards a dynamical steady state. 

\begin{figure}[ht]
    \centering
   \includegraphics[scale=0.8]{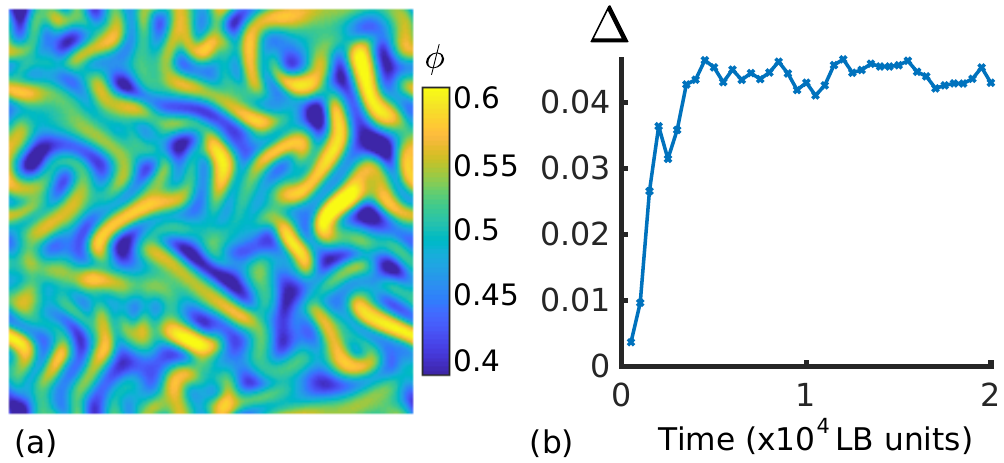}
    \caption{(color online.) {\bf Active phase ordering}: (a) Snapshot of the concentration profile in the micro-phase-separated state, color bar denotes concentration of the active phase (b) Standard deviation of the concentration\updt{,} $\Delta$, as a function of time.}
    \label{fig:ActivePhaseSeparation}
\end{figure}

\begin{figure}[ht]
    \centering
    \includegraphics[scale=0.86]{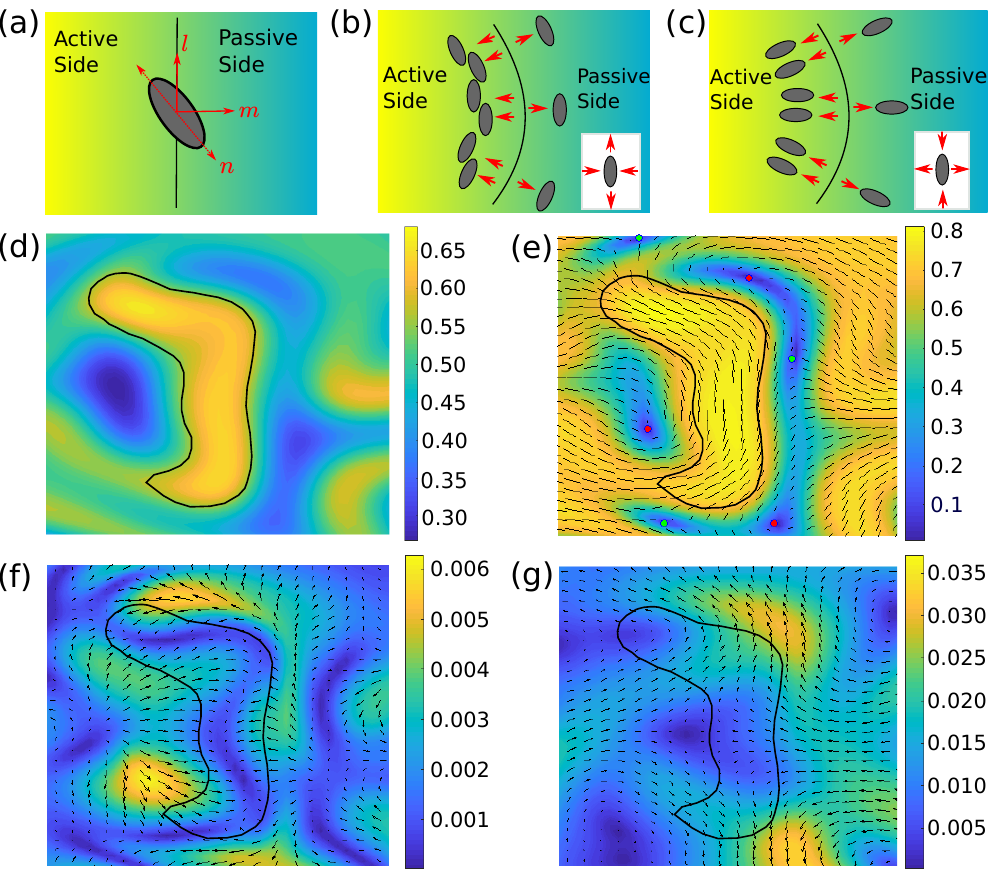}
    \caption{(color online.) {\bf Active anchoring at an interface leads to a net flow normal to the interface:} (a) Vectors introduced in the text. (b) Extensile nematics with planar active anchoring (inset: extensile flow field); (c) Contractile nematics with homeotropic active anchoring (inset: contractile flow field). In both cases the changing concentration of nematic particles leads to a net flow up the concentration gradient (red arrows). (d)-(g): Simulation snapshot of an actively phase-separated droplet. (a) Concentration field. (b) Director field, with planar active anchoring everywhere except at the tips. (c) Velocity difference between the active and passive fluid components, which points into the droplet except near $+1/2$ defects. (d) The combined flow field stretches the droplet, in a way identical to the single fluid result. }
   \label{fig:MechanismSketches}
\end{figure}

%\noindent
{\bf Mechanism.}---We emphasise that the phase separation is entirely due to the activity, and requires no passive forces.  The free energy in Eq.~(\ref{eqn:FreeEnergy}) is chosen to be a single well potential, $a>0$, which favours mixing. To explain the mechanism causing the phase ordering consider a small fluctuation in concentration. From Eq.~(\ref{eqn:totalStress}) this will lead to an active force tangential to the interface between the higher and lower activity regions %\cite{Blow2014}, 
$ \mathbf{F}^{\footnotesize{\mbox{tangential}}} = 2 \zeta  |\nabla (S\phi)| \,(\mathbf{m \cdot n}) (\mathbf{l \cdot n}) \,\mathbf{l} $ where $\mathbf{m}$ and $\mathbf{l}$ are unit vectors normal (pointing away from the more active region) and  tangential to the interface respectively, and $\mathbf{n}$ is a unit vector along the nematic director (Fig \ref{fig:MechanismSketches}(a)) \cite{Blow2014}\updt{, see also SM, Sec.~1~\cite{supp}}. 
This force sets up flows parallel to the interface.
The nematic director rotates in the flow leading to a net alignment parallel (perpendicular) to the interface in extensile (contractile) systems. This is termed active anchoring \cite{Blow2014, Ruske2022}.

 Similarly the active force normal to the interface is 
  %\begin{flalign}
      $ \mathbf{F}^{\footnotesize{\mbox{normal}}} = -\zeta  |\nabla (S \phi)| \, (2 (\mathbf{m \cdot n})^2 -1 ) \,\mathbf{m}. $
  %\end{flalign}
 In the absence of anchoring,  $F^{\footnotesize{\mbox{normal}}}$  averages to zero. However, when there is an active anchoring of the director, the force drives the active component up the concentration gradient  towards the active nematic region. Hence the relative flow between the two components
 enhances concentration fluctuations to form active regions of concentration $\phi>1/2$. The same mechanism leads to domain formation in contractile systems, as the change in the direction of the active anchoring  is compensated by the change in sign of the activity. This is illustrated schematically in Figs.~\ref{fig:MechanismSketches}(b),(c). A simulation snapshot of an active droplet is shown in Figs.~\ref{fig:MechanismSketches}~(d)-(g) .
 
 %\ju{maybe better later: The steady-state value of $\Delta$ is set by a balance between these flows and the restoring force due to the free energy. }

\begin{figure}[ht]
    \centering
    \includegraphics[scale=0.8]{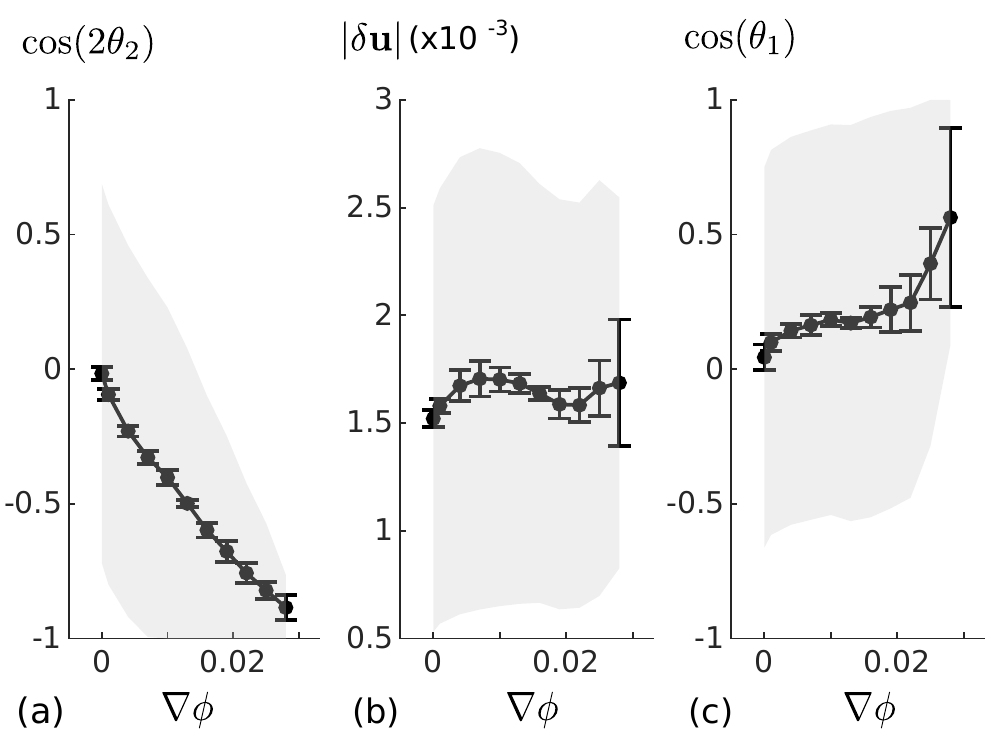}
    \caption{{\bf Flow--interface correlations:} (a) Active anchoring increases with concentration gradients. (b) Magnitude of mean slip velocity does not depend on concentration gradients. (c) Orientation of mean slip velocity points up concentration gradients. Results are for extensile activity. Shaded regions show standard deviations of observed values.  }
    %Proof of mechanism from numerical simulations of an extensile-isotropic mixture. Relevant quantities were averaged for concentration gradients within a small range of values. Shaded regions shows range of observed values. (a) Mean slip velocity does not depend on concentration gradient. (b) Mean slip orientation points along the interface for large concentration gradients. (c) Active anchoring at interfaces increases for large concentration gradients. }
    \label{fig:MechanismProof}
\end{figure}

To confirm the relevance of interfacial flows we numerically measured the dependence of physical quantities on the modulus of the concentration gradient, averaging over the whole domain. 
Projecting the nematic director field onto the  concentration gradient yields $\cos(2 \theta_2) = (\mathbf { \nabla \phi}^T \cdot\mathbf{Q} \cdot \mathbf{\nabla \phi })/(S |\nabla \phi|^2)$, where $\theta_2$ is the angle between the director field and $\nabla \phi$. This quantity tends to -1 for extensile activity confirming planar active anchoring (Fig.~\ref{fig:MechanismProof}(a)). Similarly, it approaches +1 for contractile activity corresponding to homeotropic active anchoring (Fig.~A1 in SM \cite{supp}).
The magnitude of the slip velocity $\mid \!\!\delta \mathbf{u}\!\! \mid$ is independent of the concentration gradient  $\mid \!\! \nabla \phi \!\! \mid$ (Fig.~\ref{fig:MechanismProof}(b)). However, 
 the cosine of the angle between $\delta \mathbf{u}$ and $\nabla \phi$, $\cos(\theta_1) = (\mathbf{\nabla \phi} \cdot  \delta \mathbf{u})/(|\nabla \phi||\delta \mathbf{u}|)$, increases at large $|\nabla \phi|$, confirming that the direction of the relative fluid velocity is preferentially up concentration gradients (Fig.~\ref{fig:MechanismProof}(c)).

\begin{figure}[htp]
    \centering
    \includegraphics[scale=0.7]{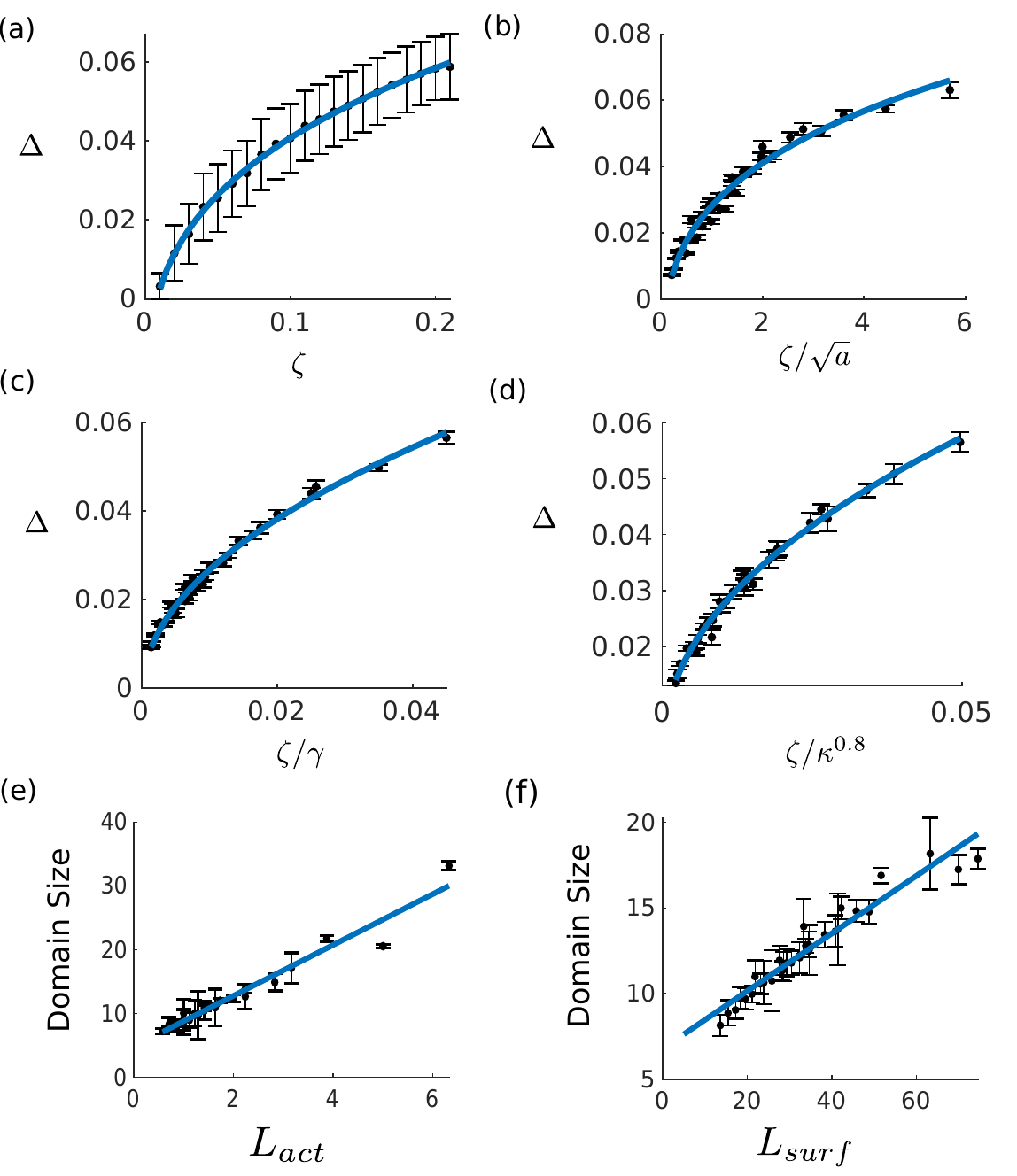}
    \caption{(color online.) {\bf Variation of concentration difference $\Delta$ {and average domain size }with model parameters:}  In all the plots above, we vary activity  $\zeta$ {$\in [0.03, 0.18]$}.(a) $\Delta$ scales non-linearly on increasing activity.  Ordering is primarily opposed by (b) bulk free energy {$a \in [0.0010, 0.0075]$}, (c) drag between the fluids {$\gamma \in [4, 22]$}, (d) surface tension {$\kappa \in [5, 25]$}. Average domain size as a function of the (e) active length-scale, $L_{\mbox{act}}${, for $K_{LC} \in [0.05, 0.40]$}. (f) Surface tension length-scale, $L_{\mbox{surf}}$, for {$\kappa \in [5, 25]$}. Blue curves are best-fit power laws for (a)-(d), and best-fit straight lines for (e), (f). Error bars show standard deviations from 40 simulation values.} 
    \label{fig:ScalingParameters}
\end{figure}

%\noindent
{\bf Varying the Model Parameters.}---We now discuss more detailed numerical results investigating how the size of the phase separated domains and the difference in concentration between the two phases, $\Delta$, depend on the model parameters. Fig.~\ref{fig:ScalingParameters}(a) shows that  $\Delta$ increases, but does not scale linearly, with the activity coefficient $\zeta$. This is because at higher activities the flow is more turbulent with many defects which reduce the strength of the active anchoring. 

The formation of concentration gradients is opposed by the bulk free energy which scales as $a$, the relative drag $\gamma$, and the surface tension $\kappa$. 
We consider the balance between the activity and each of these in turn, choosing parameters where the given relaxation mechanism is dominant. Results are presented for extensile activity but they also hold for contractile systems since the same mechanisms are at play (SM, Fig.~A2  \cite{supp}).

In Fig.~\ref{fig:ScalingParameters}(b) $\Delta$ is plotted as a function of $\zeta/a^{0.5}$ showing collapse onto a single curve. The collapse is a result of the balance between the driving force normal to the interface and the thermodynamic force which restores mixing. The former
scales as $\zeta^2$ because it depends on the strength of the anchoring and on the magnitude of the active force, both of which scale with $\zeta$.  The latter scales as $a$. 

The data also lies on a single curve when  $\Delta$ is plotted against the ratio $\zeta/\gamma$ (Fig.~\ref{fig:ScalingParameters}(c)). This shows the expected inverse relation between the activity, which creates a slipping velocity between the active and passive fluids, and the viscous drag that dampens the slip. 

A similar plot, obtained by changing the surface tension coefficient $\kappa$ at constant $a$ and $\gamma$, shows the best data collapse when $\Delta$ is plotted against $\zeta/\kappa^{0.8}$ (Fig.~\ref{fig:ScalingParameters}(d)). We do not have a simple argument for this dependence as several factors contribute. In addition to the thermodynamic penalty, an increase in surface tension weakens the activity gradient by making the interface more diffuse, and hence will alter the details of the active interfacial forces and active anchoring. %Moreover there will be a thermodynamic penalty to domain formation which changes with both the magnitude of the surface tension and the size of the domains.

The length-scale of the domains can be calculated numerically from the first zero of the spatial correlation function of the fluctuations in the concentration field. For fixed surface tension, Fig.~\ref{fig:ScalingParameters}e shows that the domain size is controlled by the usual active nematic length scale $L_{\footnotesize{\mbox{act}}} = \sqrt{{K_{LC}}/{\zeta}}$. This is, however, not the only length scale controlling the physics. The domain size can be changed by varying a second length scale, related to the surface tension, $L_{\footnotesize{\mbox{surf}}}=\sqrt{{\kappa\eta}/{\zeta}}$, as shown in Fig.~\ref{fig:ScalingParameters}f. \updt{Active droplets break up by forming bend or splay instabilities with typical lengthscale $L_{act}$, or by self-shearing instabilities \cite{Singh2019, Blow2014} with typical lengthscale $L_{surf}$. Snapshots showing formation and dissolution of a droplet are shown in SM, Fig. A3 \cite{supp}.}

%\noindent
{\bf Threshold for phase ordering.}---It is possible to obtain an exact expression for the activity threshold above which phase ordering occurs, in the limit of perfect surface anchoring, and ignoring interface curvature. \updt{The effects of curvature are discussed in Sec. 2 of the SM \cite{supp}.}

Consider a concentration gradient around $\phi = 1/2$ in the $x$-direction, given by $\nabla \phi = \mid \!\! \nabla \phi \!\!\mid \,\!\! \hat{x}$. Let the activity coefficient be $\pm |\zeta|$, where the $+$ $(-)$ sign denotes extensile (contractile) activity. Assuming strong active anchoring at the interface, this leads to a director field $Q_{xx} = \mp S_{\footnotesize{\mbox{nem}}} \updt{\sqrt\phi}, Q_{xy} = 0$.
The normal active force at the interface is then
%\begin{flalign}
   $ \mathbf{F}_{act} = \updt{(3/2)}S_{nem} \updt{\sqrt\phi} \mid \!\!\zeta \!\! \mid  \, \mid \!\!\nabla \phi \!\! \mid  \,\, \mathbf{\hat{x}}  $,
%\end{flalign}
and the restoring force driven by the free energy is 
%\begin{flalign}
  $  \mathbf{F}_{rest} =  -2a \updt{\rho\phi(1-\phi)} |\nabla \phi| \,\, \mathbf{\hat{x}} $.
%\end{flalign}
This leads to an expression for the slip velocity $\delta u_x$ in the Stokes limit:
\begin{flalign}
    \delta u_x = \frac{F_{act} + F_{rest}}{\gamma \phi (1-\phi)} = \frac{\updt{(3/2)} \mid \!\!\zeta \!\! \mid S_{nem}\updt{\sqrt\phi} - 2a \updt{\rho\phi(1-\phi)} }{\gamma \phi (1-\phi)}  \mid \!\!\nabla \phi \!\! \mid\hat{x}.
    \label{eqn:deltaU_analytics}
\end{flalign}
The concentration gradient will increase for activities greater than %\begin{flalign}
$    \mid \!\!\zeta \!\! \mid_{crit} = \frac{4 a \updt{\rho\phi(1-\phi)} }{3 \sqrt \phi S_{nem}}. $ \updt{(see SM, Fig.~A4 \cite{supp}.)}
%    \label{eqn:CritActivity}
%\end{flalign}

%The ordering mechanism will be less efficient because of surface tension and imperfect active anchoring. Therefore the activity estimated from this calculation is a lower bound on the critical activity required for active phase separation. We have checked numerically and found spontaneous domain formation near the critical activity by imposing strong thermodynamic anchoring at the interface.

%\section{}

{\bf Summary \& outlook.}--- By introducing an active two-fluid model with viscous drag between the fluids we have found that a homogeneous mixture of an active nematic fluid and a passive fluid spontaneously phase orders into regions with different concentrations of each fluid component. This is  driven dynamically by active flows set up at concentration interfaces, and is an example of liquid-liquid phase separation \cite{Hyman2014} out of thermodynamic equilibrium. 

The two-fluid model reduces to the lyotropic formulation of nematohydrodynamics that has been successful in defining active and passive regions within a simulation, in the limit of infinite viscous drag between the fluids. However, because the lyotropic approach uses a single velocity field for both phases \cite{Blow2014, Giomi2014}, it cannot account for relative flows between, or compressibility of, the two components. 

%Two-fluid models \cite{Drew1983, stewart1984, Swift1996} overcome this limitation, and have been used to study the cytoplasm-cytoskeleton system \cite{Levine2009, Wolgemuth2020, Mogilner2018}, and the motion of cells \cite{Alt1999, Cogan2010}. Our model reduces to the  lyotropic model in the limit of infinite viscous drag between the fluids, and extends the lyotropic model to accurately study multi-phase flows in biological systems.

%Our results predict the spontaneous formation of active droplets and active-passive interfaces. Active droplets are stable, and can move and divide  \cite{Joanny2012,Giomi2014}. An active-passive interface is stable \cite{Coelho2019, Soni2019}, supports propagating modes \cite{Wysocki2016, Patteson2018}, and gives rise to interfacial fluctuations and active waves \cite{Adkins2022}. 

Active nematics have been observed to preferentially adhere to a substrate or container wall, a phenomenon known as active wetting \cite{Adkins2022, Perez-Gonzalez2019, Alert2020, Banerjee2019}.
A straightforward extension of our argument shows that for an initially homogeneous
active-isotropic mixture, the normal force generated at a boundary would cause wetting of the boundary by the
active nematic for both extensile and contractile activity, even in the absence of surface thermodynamic forces. The wetting would then be enhanced by any thermodynamic planar (normal) anchoring for extensile (contractile) activity.

% if the thermodynamic anchoring at the boundary is the same sign
%as active anchoring - planar (homeotropic) anchoring for extensile (contractile) active nematics.
%%On the other hand, the same effect would cause active fluid dewetting if the thermodynamic anchoring
%at the boundary is of opposite type to active anchoring - homeotropic (planar) anchoring
%for extensile (contractile) active nematics.

In future work it will be interesting to study mixtures where both the fluids are active. In addition to cell sorting this is relevant to the growth and dynamics of bacterial and algal colonies 
which contain more than one species \cite{RAMANAN2016}.  Moreover, the relaxation of the incompressibility constraint will allow the study of how compressible flows affect defect behaviour, and whether they can stabilize structures such as asters or the cellular lumen.

\vspace{5mm}

%\section*{Acknowledgements}
\begin{acknowledgements}
{\bf Acknowledgements.}---We thank L. J. Ruske and I. Hadjifrangiskou for valuable discussions. SB acknowledges support from the Rhodes Trust and the Crewe Graduate Scholarship. %acknowledge useful discussions with 
\end{acknowledgements}

%\bibliography{refs}


\end{document}